\def\year{2015}

\documentclass[letterpaper]{article}
\usepackage{aaai}
\usepackage{times}
\usepackage{helvet}
\usepackage{courier}
\usepackage{multirow}
\usepackage{balance}  
\usepackage{graphics} 
\usepackage{times}    
\usepackage{url}      
\usepackage{balance}  
\usepackage{graphicx}
\usepackage{subfig}
\usepackage{epstopdf}
\usepackage{rotating}

\usepackage[utf8]{inputenc}

\begin{document}

\title{Smelly Maps: The Digital Life of Urban Smellscapes}

\author{Daniele Quercia\\University of Cambridge\\\textit{dquercia@acm.org} \And
Rossano Schifanella\\University of Torino\\\textit{schifane@di.unito.it} \And
Luca Maria Aiello\\Yahoo Labs\\\textit{alucca@yahoo-inc.com} \And
Kate McLean\\Royal College of Art \& CCCU\\\textit{mcleankate@mac.com} }

\maketitle

\begin{abstract}
\begin{quote}
Smell has a huge influence over how we perceive places. Despite its importance, smell has been crucially overlooked by urban planners and scientists alike, not least because it is difficult to record and analyze at scale. One of the authors of this paper has ventured out in the urban world and conducted ``smellwalks'' in a variety of cities: participants were exposed to a range of different smellscapes and asked to record their experiences. As a result, smell-related words have been collected and classified, creating the first dictionary for urban smell.  Here we  explore the possibility of using social media data to reliably map the smells of entire cities. To this end, for both Barcelona and London, we collect geo-referenced picture tags from Flickr and Instagram, and geo-referenced tweets from Twitter. We match those tags and tweets with the words in the smell dictionary. We find that smell-related words are best classified in ten categories. We also find that specific categories (e.g., industry, transport, cleaning) correlate with governmental air quality indicators, adding validity to our study.
\end{quote}
\end{abstract}

\section{Introduction}
\label{sec:introduction}

Smells impact our behavior, attitudes and health. Street food markets, for example, have dramatically changed the way we perceive entire streets of global cities.

Despite its importance (which we will detail in Section~\ref{sect:whysmell}), smell has been crucially overlooked (Section~\ref{ref:related}). City planners are mostly concerned with managing and controlling bad odors. Scientists have focused on the negative research aspects of smell as well: they have studied air pollution characteristics (often called `environmental stressors') rather than the more general concept of smell. As a result, the methodological tools at the disposal of researchers and practitioners  are quite limited. Smell is simply hard to capture.

To enrich the urban smell toolkit, we here explore the possibility of using social media data to reliably map the smells of entire cities. In so doing, we make the following contributions:

\begin{itemize}

\item One of the authors of this paper ventured out in the urban world and conducted ``smellwalks'' around seven cities in UK, Europe, and USA (Section~\ref{sec:smellwalks}). Locals were asked to walk around their city, identify distinct odors, and take notes. Smell descriptors are taken verbatim from the smell walkers' original hand-written notes. As a result of those sensory walks, smell-related words were recorded and classified, resulting in the first urban smell dictionary, which we will make publicly available to the research community.

\item For the cities of Barcelona and London, we collected  geo-referenced tags from about 530K Flickr pictures and 35K  Instagram photos, and 113K geo-referenced tweets from Twitter (Section~\ref{ref:dataset:socialmedia}). We matched those tags and tweets with the words in the smell dictionary.

\item We found that smell-related words are best classified in ten categories (Section~\ref{ref:methodology:clustering}). Our classification, generated automatically from social media, is very similar to classification systems obtained manually as a result of field research.

\item We also found that specific categories (e.g., industry, transport, cleaning) correlate with governmental air quality indicators (Section~\ref{ref:results}), and that speaks to the validity of our study. Finally, we show that, using social media data, we are able to capture not only a city's dominant smells (base smell notes) but also localized ones (mid-level smell notes). 
\end{itemize}

These results open up new opportunities (Section~\ref{sec:discussion}). Our ultimate goal is to open up a new stream of research that celebrates the positive role that smell has to play in city life.

\section{Why Smell}
\label{sect:whysmell}

Our daily urban experiences are the product of our perceptions and senses~\cite{quercia14shortest,quercia15walk}, yet the complete sensorial range is  strikingly absent from urban studies. Sight has been historically privileged over the other senses. In the early sixties, Jane Jacobs stressed the importance of visual order in the city~\cite{jacobs1961death}, and Kevin Lynch focused on the visual dimensions of urban design~\cite{lynch1960}.  When odor is mentioned in the built environment literature, it is generally in negative terms. We do not have to go that far back in time to find the first positive reference to smell by a celebrated architect: in 2005, Juhani Uolevi Pallasmaa  briefly highlighted smell in the second part of his well-known book `The Eyes of the Skin'~\cite{pallasmaa2012eyes}.

The problem is that not knowing what smells exist in cities may result in:

\begin{description}

\item \emph{Partial views of the collective image of the city.} Good cities are those that have been built and maintained in a way that they are imageable, i.e., that their mental maps are clear and economical in terms of mental effort~\cite{lynch1960}.  Urban studies have explored how imageability is affected by the ease with which people memorize visual images of the city.  Yet, memory is affected not only by what we see but also by what we smell. Smell and long-term memory are closely related and, more importantly, odor associations are  retained for much longer time periods than visual images~\cite{engen1991odor}. Research into how we build the collective image of the city has overlooked the significant role that smell plays in our urban well being.

\item \emph{The proliferation of clone towns.}  Odors contribute to a place's identity. Place identity odors are often overlooked by city planning professionals who either do not notice them, or do not attribute any value to them. This results into the proliferation of homogenized, sterile and controlled areas that have ``an alienating sense of placeness.''~\cite{drobnick2006smell}. Because of globalization, the typical UK town or city has become a `clone town': ``a place that has had the individuality of its high street shops replaced by a monochrome strip of global and national chains''.  Clone towns can only  offer homogenized olfactory environments~\cite{reynolds2009guerrilla}.

\item  \emph{Reinforcing socio-economic boundaries.} Odors contribute to the construction of a place's socio-economic identity. The greasy odors coming from fast food restaurants are often associated with rundown areas and with the evening economy. \cite{Macdonald2007251} found a significant positive relationship between the location of the four most popular fast food chains  in UK and neighborhood socio-economic deprivation. Smells that provide insights into the social life of cities are used as an invisible marker in reinforcing socio-economic boundaries. If we do not know what smells exist, we are likely to reinforce those boundaries without even knowing it.

\end{description}

\section{Related work}\label{ref:related}

People are able to  detect up to 1 trillion smells~\cite{bushdid14}. Despite that, there are limited maps of this potentially vast urban smellscape. One reason is that smell is problematic to record, to analyse and to depict visually. Here we review a variety of methodological approaches for recording urban smells.

\mbox{ } \\
\textbf{Recording odors with devices.}  Olfactometers have been used to collect information about distinct odor molecules. They look like `nose trumpets'  and capture four main aspects: odor character, odor intensity,  duration, and frequency. Public agencies usually use them to verify complaints about odor nuisances. Other smell recording technologies include a head-space `smell camera'. This device traps volatile odor molecules in a vacuum and  is able to capture permanent (i.e., non fleeting) smells. 

\mbox{ } \\
\noindent \textbf{Recording odors with the Web.} Online participatory mapping allows web users to annotate pre-designed base maps with odor markers~\cite{victoria2013}. This method promises to be scalable, but engaging enough people to participate is hard. 

\mbox{ } \\
\textbf{Recording odors with sensory walks.}  Social science research has increasingly used the methodology of sensory walks. The earliest example of a sensewalk was undertaken in 1967 by Southworth with a focus on the sonic environments of cities~\cite{southworth1967sonic}. During the soundwalk, participants were involved in increasingly focused tasks of attentive listening.   Urban smellwalks are similarly designed. They consist of  incrementally broadening experiences to cover the wider olfactory environment. A few multi-sensory researchers have engaged in such walks. 

In Vienna in 2011 the philosopher Madalina Diaconu ran a project exploring the meanings and associations of the tactile and olfactory qualities of the city researching through the noses of a group of students~\cite{diaconu2011senses}. Meanwhile, Victoria Henshaw conducted her smellwalks in Doncaster, England~\cite{victoria2013}. Most of our work is based upon her research findings. The problem is that the sensory walk methodology collects fine-grained data but is not scalable. To see why, consider that an individual walk typically took Henshaw three non-continuous months, involved six participants, and covered approximately 160 $km^2$.  \\

To sum up, previous odor collection methodologies are not scalable. Web-based methodologies would be only  under the unrealistic assumption of massive public engagement. By contrast, sensory walks successfully engage one individual at the time but, when carried out over several years, they only result in data about limited geographic areas. We thus need a new way of collecting odor information at scale without requiring a massive public engagement. 

\section{Methodology} \label{sec:methodology}

This work proposes  a new way of doing so from data implicitly generated by social media users. The idea is to search for smell-related words on geo-referenced social media content. To do so, we need those words and the content itself, both of which are described in the following section. 

\subsection{Urban Smell Dictionary}
\label{sec:smellwalks}

When attempting to control and enforce odor law and policies, city authorities face the well-known difficulty of recording, measuring, describing, and classifying odors. Smells do not lend themselves easily to quantitative measurement. However, scientists have long attempted to propose a unified odor categorization system. 

\mbox{ } \\
\textbf{Aristotelean Classification.} Aristotle, for example, divided odours into six separate classes, later amended by Linnaeus in 1756 to seven; aromatic, fragrant, alliaceous (garlic), ambrosial (musky), hircinous (goaty), repulsive and nauseous. 

\mbox{ } \\
\textbf{UCLA Odor Classification.} One issue is that, to describe smells, people often name potential sources of smell rather than actual odors. They, for example, use terms such as sulphurous, eggy, floral, earthy, or nutty. To partly fix this problem,~\cite{curren12} developed an urban odor descriptor wheel that includes the words people use to describe specific odors (e.g., grease) along with their chemical names (e.g, 2-meythyl isobomeol).

\mbox{ } \\
\textbf{Henshaw's classification.} For her PhD, Victoria Henshaw set out to document odors present in the city of Doncaster. She did so by conducting a number of smellwalks. These smellwalks followed a  pre-planned set of routes. Each route was designed in a way that provided exposure to a range of different smellscapes. The route included a set of stopping points (e.g., mixed-used developments, busy bus routes, ethnically diverse residential areas, business areas, markets). At each stopping point area, a range of questions regarding the smells that the participants detected were asked. Those questions invited insights not only about annoyance and disturbance (as it is usually done) but also about positive perceptions of smell. After each smellwalk, the weather, temperature, time, and activities taking place were recorded. The different walks were carried out in periods of cold weather (e.g., January-March) and of warmer weather (e.g., April-July), on weekdays and weekends, and at different times of the day from 7am to 8pm.  Henshaw also combined her insights with those offered by the Vivacity2020 Project that investigates urban environmental quality. This combination resulted into a  classification of urban odors along 11 types: traffic emissions, industrial odors, food and beverages, tobacco smoke, cleaning materials, synthetic odors, waste, people and animals, odors of nature, building materials, and non-food items.

\begin{figure*}[t!]
\begin{center}
\includegraphics[width=0.7\textwidth]{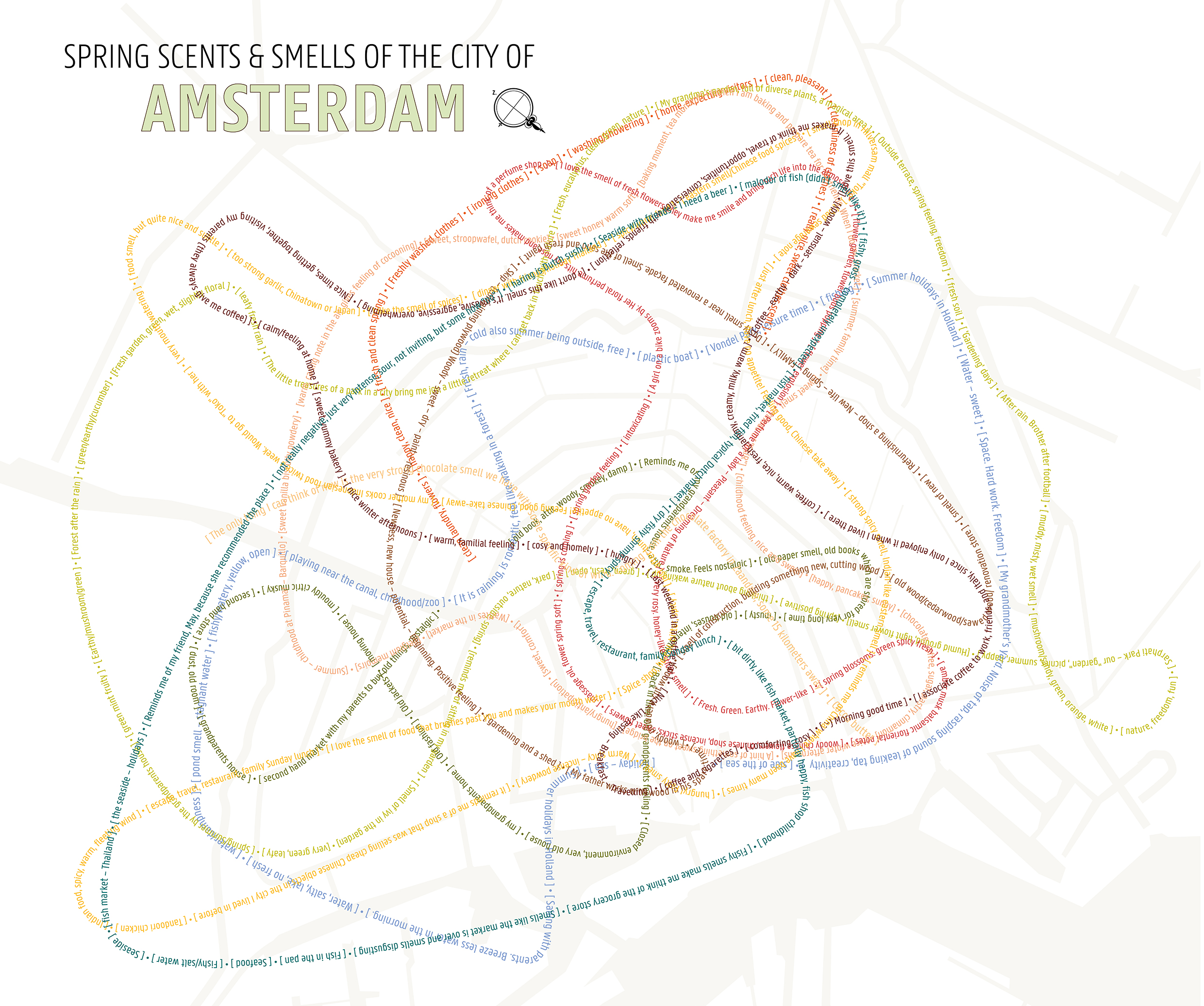}
\caption{Visualization of the hand-written annotations taken by participants of the smellwalk in Amsterdam.}
\label{fig:kate}
\vspace{-2mm}
\end{center}
\end{figure*}

\begin{table}[t!]
\centering
\begin{tabular}{l|c|c}
\hline
City & \#participants &  \#smells \\
\hline
Amsterdam & 44 & 650 \\
Pamplona & 58 & 374 \\
Glasgow & 20 & 55 \\
Edinburgh & 10 & 40 \\
Newport & 30 & 80 \\
Paris & 10 & 25 \\
New York & 20 & 43 \\
 (Brooklyn+Greenwich Village) &  & \\
\hline
\end{tabular}
\caption{The set of smellwalks whose data has been used to integrate previous odor classifications.
For example, in Amsterdam, the smellwalks involved  44 local residents over 4 days in April 2013 and resulted in the collection of 650 smell perceptions, which include background smells, episodic and unexpected occurrences. }
\vspace{-1em}
\label{table:walks}
\end{table}

\mbox{ } \\
\textbf{Additional Smellwalks.} One of the authors of this paper complemented Henshaw's classification by conducting  smellwalks in other cities across the UK, Europe, and USA (Table~\ref{table:walks}). These walks mainly involved local people. Participants identified distinct odors and recorded their location, description, expectation, intensity, personal association and hedonic scale. Smell descriptors are taken verbatim from the original hand-written notes. Figure~\ref{fig:kate} overlays some notes gathered at the Amsterdam's smellwalk on the city map. 

\mbox{ } \\
\noindent \textbf{Comprehensive smell dictionary.} To build a smell dictionary, we hand-code  the previously discussed literature and the hand-written notes from the smellwalks. Specifically, we use \emph{line-by-line} coding to generate a set of words conceptually associated with smell. Three annotators independently generated a list of words that relate to olfactory perceptions. We then combine the three lists using the most conservative approach; we take their \emph{intersection} (rather than union). We double-checked the resulting list removing potentially ambiguous tags (e.g., the word `orange' can refer to a fruit, a color, or a smell). The result of the processes explained above is the first urban smell dictionary containing some 285 English terms\footnote{To support future work in the area, the \url{link} to the urban dictionary will be provided in the camera-ready version.}. Since our analysis considers not only London but also Barcelona, we also manually translated the terms into Spanish. By visual inspection, one sees that all the words in the dictionary are related to the domain of smell. However, by no means, do they represent an exhaustive list. Therefore,  it is not clear whether we will observe any relationship between the presence of specific smell words in a place and the actual smell of the place.

\subsection{Social Media Data}
\label{ref:dataset:socialmedia}

\begin{table*}[t!]
\centering
\begin{tabular}{c|cccc|cccc}
& \multicolumn{4}{c|}{London} &\multicolumn{4}{c}{Barcelona} \\
          & Users & Items & Smell words & Street segments & Users & Items & Smell words & Street segments\\
\hline
Flickr    & 28.381 &  454.484 & 593.602  & 27.232 & 8.366  & 74.381 & 102.876 & 14.952  \\
Instagram &  5.509 &   30.432 &  58.522  & 11.654 &  1.513 &  5.637 &  11.314 &  5.380  \\
Twitter   & 16.214 &  109.269 & 125.137  &  9.373 &    816 &  3.915 &   4.670 &  2.245  \\
\hline
\end{tabular}
\caption{Dataset statistics for Flickr and Instagram photos and for tweets.}
\vspace{-1em}
\label{tab:datasets}
\end{table*}

Having defined the smell words, our next step is to gather social media data against which those words are matched.

\mbox{} \\ 
\textbf{Flickr.} Out of the set of all the public geo-referenced Flickr pictures, we selected a random sample of 17M public photos taken within the bounding boxes of London and Barcelona. For each picture, we collected the anonymized \textit{owner} identifier and the free-text \textit{tags} attached to the photo by the owner. 

\mbox{ } \\
\textbf{Instagram.} To obtain a sizable sample of Instagram pictures, we collected data for a random set of 5.1M  users whose accounts were public. We collected all of their ``feeds'' for a three-year period between December 2011 to December 2014. The collection resulted in about 154M images and videos along with their meta data including \textit{hashtags}, \textit{captions}, and \textit{geo-references}. Using the picture geo-location, we selected photos taken in London and Barcelona, for a total of 436K images.

\mbox{ } \\
\textbf{Twitter.} We gather geo-referenced \textit{tweets}. Using the Twitter API, we collected 5.3M tweets  during year 2010 and from October 2013 to February 2014. Out of those, we selected the 1.7M geo-referenced tweets in London and Barcelona after filtering out retweets and direct replies.

\subsection{Urban Smell Classification}\label{ref:methodology:clustering}

We then textually parsed our geo-referenced items (which are tags in Flickr, hashtags and captions in Instagram, and tweets in Twitter) and searched for exact matches with the dictionary words. Table~\ref{tab:datasets} summarizes the size our datasets together with the total number of matched smell words. To verify whether those words matched pictures that actually related to smell, we manually checked 100 random Flickr pictures and found that $85\%$ of the pictures did so. 

\begin{figure}[t!]
\begin{center}
\includegraphics[width=0.99\columnwidth]{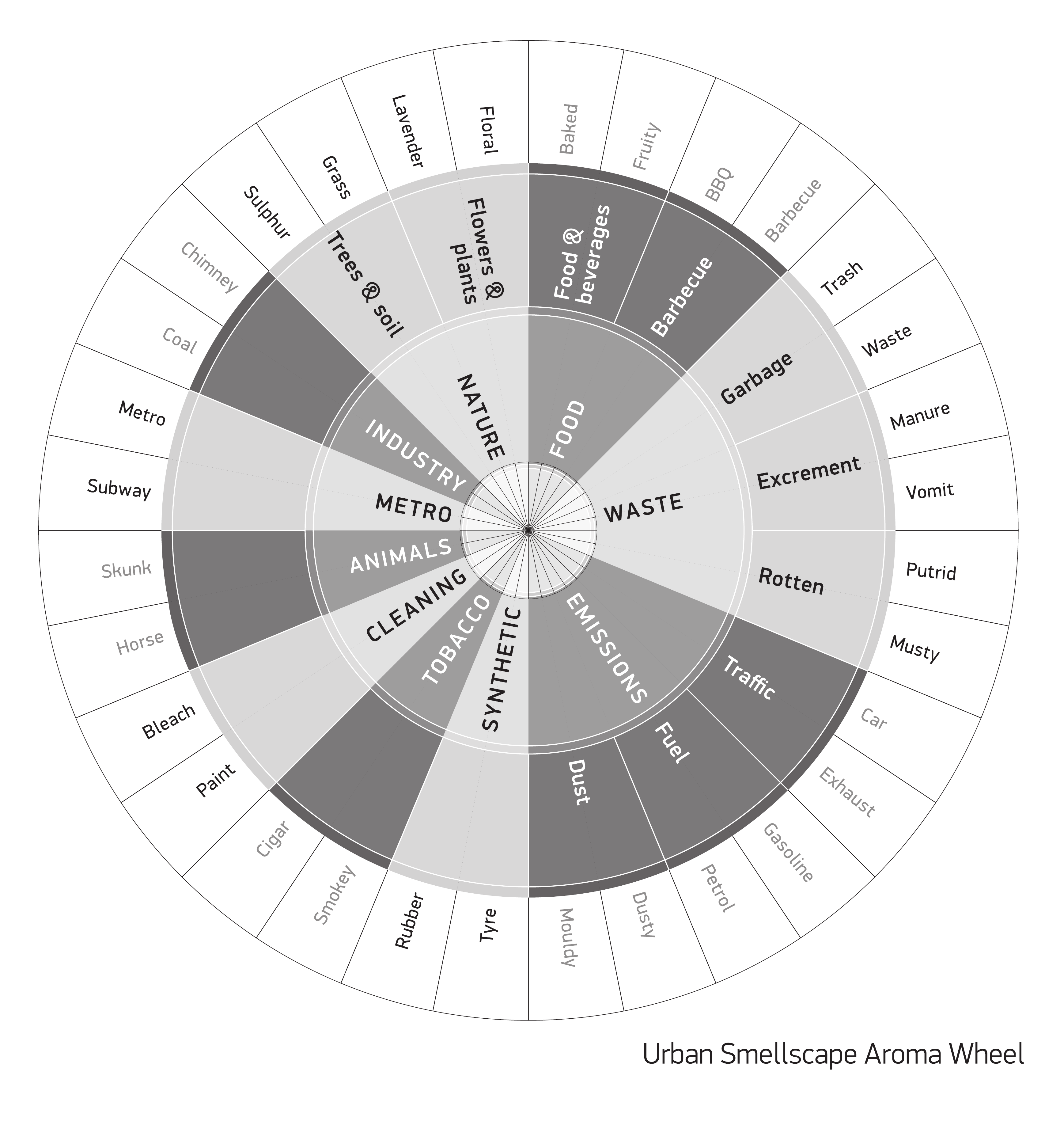}
\caption{Urban smellscape taxonomy. Top-level categories are in the inner circle; second-level categories, when available, are in the outer ring; and examples of words are in the outermost ring.}
\label{fig:taxonomy}
\vspace{-5mm}
\end{center}
\end{figure}

The next stage was to create a structure for a large and apparently unrelated dataset of smell words through a system of classification. We first built a co-occurrence network where nodes are smell words and undirected edges are weighted with the number of times the two words co-occur in the same Flickr pictures  as tags (Flickr is the dataset containing the highest number of smell words). We built this co-occurrence network because the  semantic relatedness among words naturally emerges from the  network's \textit{community structure}: semantically related nodes are those that are highly clustered together and weakly connected to the rest of the network. To determine the community structure, we could use any of the literally thousands of different community detection algorithms that have been developed in the last decade~\cite{fortunato10community}. None of them always returns the ``best'' clustering. However, since Infomap has shown very good performance across several benchmarks~\cite{fortunato10community}, we opt for using it to obtain the initial partition of our network~\cite{rosvall08maps}. This partition results in many clusters containing semantically-related words, but it also results in some clusters that are simply too big to possibly be semantically homogeneous. To further split those clusters, we iteratively apply the community detection algorithm by Blondel \emph{et al.}~(\citeyear{blondel08fast}), which has been found to be the second best performing algorithm~\cite{fortunato10community}. This algorithm stops when no node switch between communities  increases the overall modularity\footnote{If one were to apply  Blondel's right from the start, the resulting clusters are less coherent than those produced by our approach.}. The result of those two steps is the grouping  of smell words in hierarchical categories.  Since a few partitions of words might be too fine-grained, we manually double-check whether this is case and, if so, we merge all those sub-communities that are under the same hierarchical partition and that contain strongly-related smell words.  

Figure~\ref{fig:taxonomy} sketches the resulting classification. It has  ten main categories, each of which has a hierarchical structure with variable depth from 0 to 3. For brevity, the figure reports only the first level. 

This classification is of good quality because of two main reasons. First, despite spontaneously emerging from  word co-occurrences, our classification strikingly resembles Henshaw's. The only difference between the two is that ours has the category \textit{metro}\footnote{This category's words refer to public transportation facilities, and might well be a Flickr-specific artifact: London subway stations have long been of photographic interest and, as a result, might be overrepresented on Flickr.}. 

\begin{figure}[t!]
\begin{center}
\includegraphics[width=0.99\columnwidth]{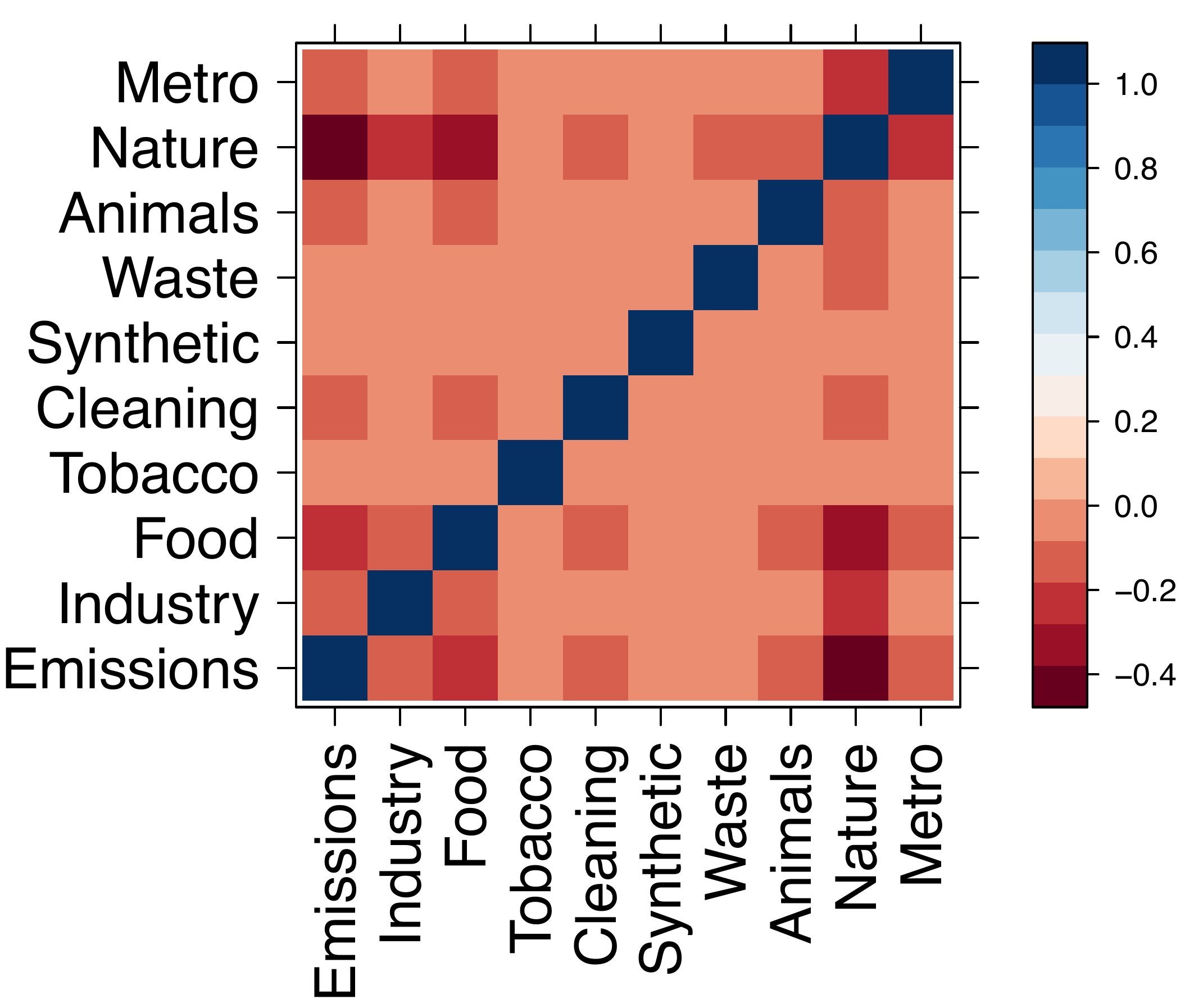}
\caption{Pairwise correlations between presence of smell categories at street level in London}
\label{fig:cross_correlation}
\vspace{-5mm}
\end{center}
\end{figure}

Second, our smell categories are ecological valid and are mostly orthogonal to each other. They are ecologically valid because, later on, will study their distribution across London streets and see that they behave as expected (Section~\ref{ref:results}). For now, just consider the pairwise correlations between the presence of a category and that of another category at street level in London (Figure~\ref{fig:cross_correlation}). By looking, for example, at the last row  (that of the category \textit{emissions}), we see that the most complementary category to it is \emph{nature}: this means that gas emissions are rarely found where greenery is found (and vice-versa). More importantly, the whole correlation matrix suggests that the vast majority of category pairs show no correlation, and that is good news because it implies that our categories are orthogonal and, as such, the clustering algorithms have done a good job.

\subsection{Air quality of streets}

The olfactory experience of a city is inevitably influenced also by the quality of the air, measured by the amount of pollutants that are emitted in the atmosphere by several human activities. It is useful to clarify the differences between air pollutants and odors. Air pollutants are chemicals that, when released into the air, pose potential harm to human and environmental health. These chemicals may or may not be detected through the human senses~\cite{McGinley00}. Some air pollutants have odors (e.g., benzene has a sickly sweet odor) while others, such as carbon monoxide, cannot be detected through the senses of smell. Air pollution is the world's largest single environmental health risk, being the cause of one in eight of the total premature global deaths, according to the World's Health Organization\footnote{\scriptsize\url{http://www.who.int/mediacentre/news/releases/2014/air-pollution/en/}}. A few pollutants are systematically measured in cities:

\begin{description}
	\item $CO$. Carbon Monoxide is a colorless, odorless poisonous gas produced by the incomplete or inefficient combustion of fuel. The gas affects the blood's transport of oxygen around the body and to the heart.
	\item $NO_2$. Nitrogen oxides are formed during high-temperature combustion processes from the oxidation of nitrogen in the air. It is a noxious gas with serious health implications: eye irritation,  irritation of the respiratory system, and shortness of breath.
	\item $O_3$. Ozone is not directly emitted, but is formed by a complex set of chemical reactions. Like $NO_2$, high levels of $O_3$ can irritate and inflame the lungs, possibly causing migraine and coughing.
	\item $PM10$, $PM2.5$. These are coarse particles ($PM10$) and fine particles ($PM2.5$) that are linked to lung cancer and asthma.  They are named for the size, in microns, of the particles. Particulate matter smaller than about $10\mu m$ ($PM10$) can settle in the bronchi and lungs and cause health problems. $PM2.5$ is the smallest and most dangerous sort of particulate matter (particles less than $2.5\mu m$ in diameter) and can enter deep into the respiratory system.
	\item $SO_2$. Sulphur dioxide  results from burning coal or oil, and it makes buildings crumble and lungs sting. 
\end{description}

It is not easy to assess health risks by comparing pollutants, not least because pollutants come at different concentration levels at any point in time. To ease comparison, the air-quality index ($AQI$) has been introduced. This rescales the concentrations of a given pollutant in the range from 1 (low risk) to 10+ (``hazardous'' for all).

In London, we collect air quality indicators as the  public API provided by the environmental research group at King's College London\footnote{\scriptsize\url{http://api.erg.kcl.ac.uk}} allows us to do. More specifically, we are able to collect $AQI$ values directly from air quality \emph{tracking stations}: $AQI$ values for $NO_2$ from 90 stations, for $PM10$ from 77, for $O_3$ from 25, for $PM2.5$ from 20, and for $SO_2$ from 13. From those numbers, it is clear that not all pollutants are measured by all tracking stations. We are also able to collect the \emph{predicted} pollutant concentration values of $NO_2$, $PM10$, and $PM2.5$ for every single street. These values are accurately estimated by  advanced models of dispersion assessments~\cite{beevers13air}. In a similar way, in Barcelona, we gather the predicted $NO_2$ pollution concentration for every street. The values are estimated by the regression models developed within the ESCAPE project~\cite{eeften12development,beelen13development}.

\subsection{Mapping data onto streets}

All this social media and air quality data now needs to be mapped. A street segment is the unit with  the finest-grained spatial resolution that is common to all our sets of data. A segment is a street's portion between two road intersections. We gathered street segment data for Central London (36.755 segments) and Barcelona (44.044 segments) from  OpenStreetMap (OSM) (a global group of volunteers who maintain free crowdsourced online maps).  After mapping our social media data onto  street segments, each segment ends up being characterized by the presence or absence of words (i.e., of smell categories) within it. Since geo-referencing comes with positing errors, we buffer each street's polyline with an area of 22.5 meters on each side. This means that  data mistakenly positioned within that buffer area is still considered part of the segment.

\section{Results}\label{ref:results}

By having our social media and air quality data mapped onto street segments, we are now able to study how those two datasets  are statistically related.  However, before doing so, we need to introduce the concept of odor notes. 

\subsection{Odor Notes} 

To best interpret our results, we should think about the different levels at which place odors can be considered. To ease illustration, we resort to Victoria Henshaw's analogy taken from the perfume industry.  When a new perfume is created,  different top, middle, and base note ingredients are combined to make the new fragrance. Those notes differ in terms of their tenacity. Top notes are those perceived immediately (e.g., citrus fruits, aromatic herbs) and, since they are intense, they are also volatile and evaporate quickly. By contrast, base notes are those adding depth and stay on the skin for hours (e.g., wood, moss, amber, and vanilla). Middle notes sit somewhere in between (e.g., flowers, spices, berries). The urban smellscape is similarly composed: 
\begin{description}
\item \emph{Base notes.} The macro-level base notes for the urban smellscape are those that are  likely smelled by a city's first-timer visitors. That is because known odors are unconsciously  processed by people, while only unfamiliar or strong odors are brought to people's attention (as potential threats or sources of pleasure). As a result, residents are not likely to pay attention to their city's base notes, while visitors would be able to consciously process them.

\item \emph{Mid-level notes.} As one moves through the city, the base notes blend with dominant smells that are localized in specific areas (e.g., factories, fish markets).

\item \emph{High notes.} Finally, the micro-level high notes are short-lived odors (e.g., goods from a leather shop). These are emitted in points that are very localized  in space and time.
\end{description}

With our analysis based upon social media, we are after the detection of base notes (uniformly distributed across the city) and mid-level notes (localized in specific areas of the city). High notes are likely to go undetected because of data sparsity and because of our spatial unit of analysis being a street segment. 

\subsection{Base Notes of Urban Smell}

\begin{figure}[t!]
\begin{center}
\includegraphics[width=0.99\columnwidth]{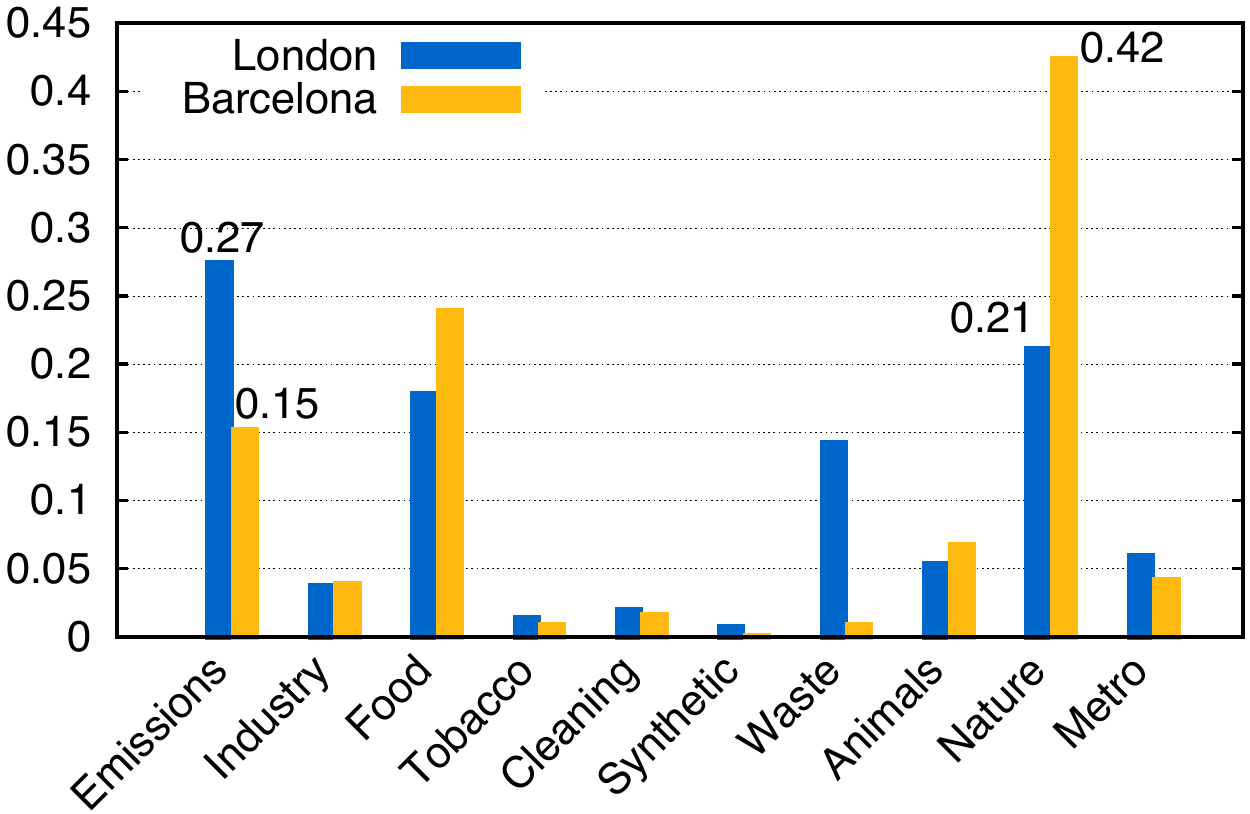}
\caption{Distribution of smell categories in London and Barcelona}
\label{fig:smell_distr}
\vspace{-5mm}
\end{center}
\end{figure}

To capture the base notes of the urban smellscape, we consider our 10-category classification of urban smells (Figure~\ref{fig:taxonomy}) and compute the fraction of Flickr tags that match each of them. This gives us a high-level olfactory footprint of the city. Barcelona is predominantly characterized by smells related to food and nature, while London is characterized by smells related to, traffic emissions and waste (Figure~\ref{fig:smell_distr}). As The Economist puts it: ``In 2013 the annual mean concentration of $NO_2$ on Oxford street [in London] was one of the highest levels found anywhere in Europe.''~\cite{economist15big}. The predominance of traffic over nature comes as no surprise since smells of traffic pollution have been found to overlay and mask more subtle smells. Air pollutants also have been found to reduce the ability of floral scent trails to travel through air.

However, critics might rightly say that the predominance of certain smell words over others might well come from data artifacts and might not  reflect the actual street odors experienced on the streets. To partly counter that argument, we test weather air quality conditions are related to the presence of specific smell words by answering this question: 

\mbox{ } \\
\textbf{Q1.} \textit{Do air quality indicators correlate with specific smell categories as expected?}
To answer that, we compute the fraction of each segment's tags that belong to a given smell category. We do that for segments having at least  30 geo-referenced smell tags to avoid data sparsity.  For each segment, we thus have a 10-element \emph{smell vector} (as there are 10 smell categories) and a set of \emph{air quality vectors} reflecting the pollutant concentrations on the segment. We need to compute the correlation between the smell vectors and the air quality vectors. However, when high spatial auto-correlation occurs, traditional metrics of correlation such as Pearson  require independent observations and cannot thus be directly applied. To overcome this problem, we used a statistical method introduced by Clifford \emph{et al.}~\cite{clifford89assessing}. This approach addresses the ``redundant, or duplicated, information contained in georeferenced data''~\cite{griffith11nonstandard} -- the effect of spatial auto-correlation -- through the calculation of a reduced  sample size. After doing so, we find our hypothesized relationships to hold: pollutant concentrations are positively correlated with the  category of (traffic) \emph{emissions} in the smell vectors ($r=0.47$ in London and $r=0.29$ in Barcelona for $NO_2$, $p < 0.001$) and are negatively correlated with the category of \emph{nature} ($r=-0.33$ in London, $r=-0.35$ in Barcelona for $NO_2$, $p < 0.001$). One might wonder whether those results change across social media platforms. 

\begin{figure}[t!]
\begin{center}
\includegraphics[width=0.95\columnwidth]{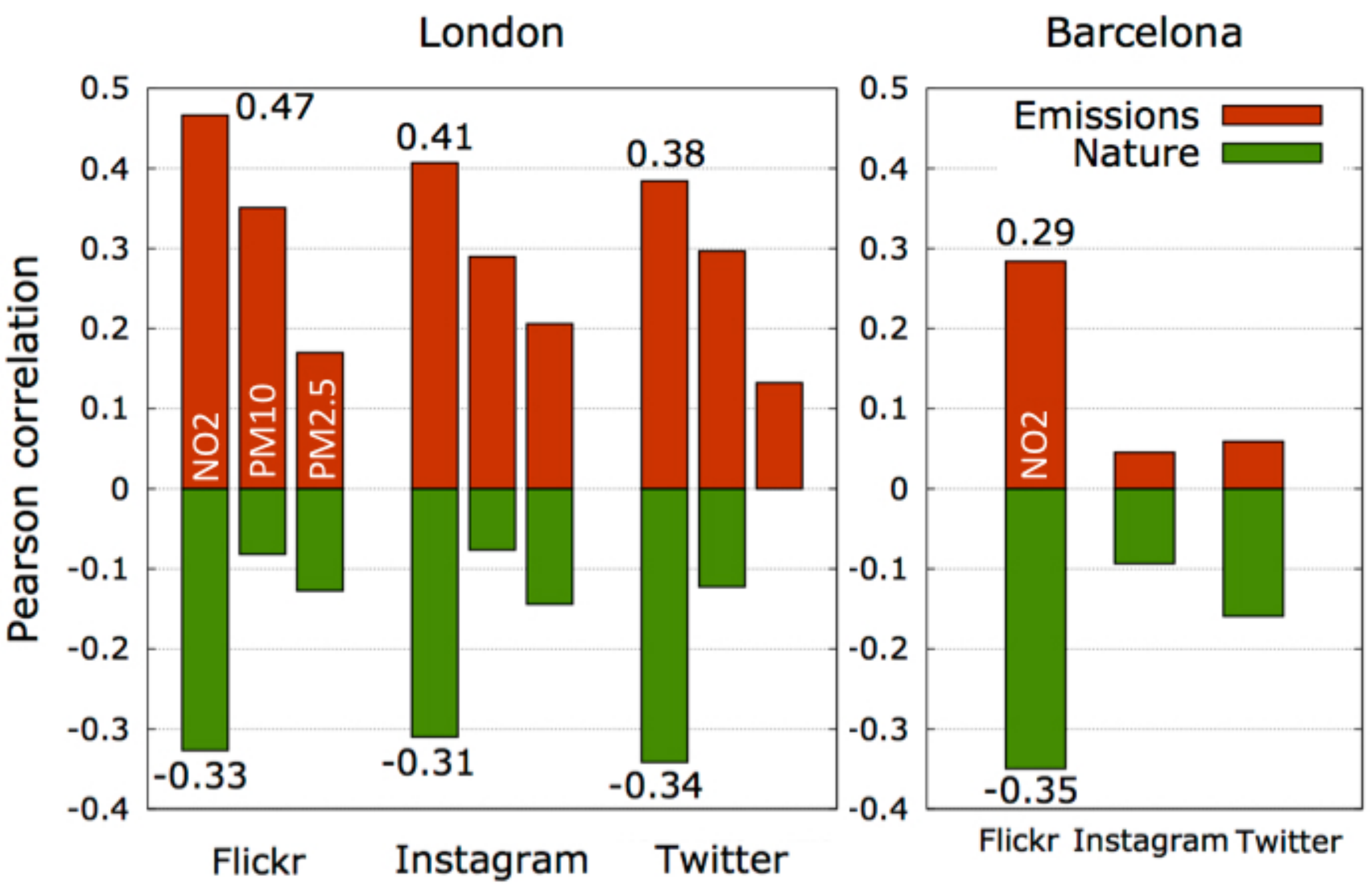}
\caption{Correlations between smell and pollutants, across social platforms}
\label{fig:correlations}
\vspace{-5mm}
\end{center}
\end{figure}

\begin{figure}[t!]
\begin{center}
\includegraphics[width=0.99\columnwidth]{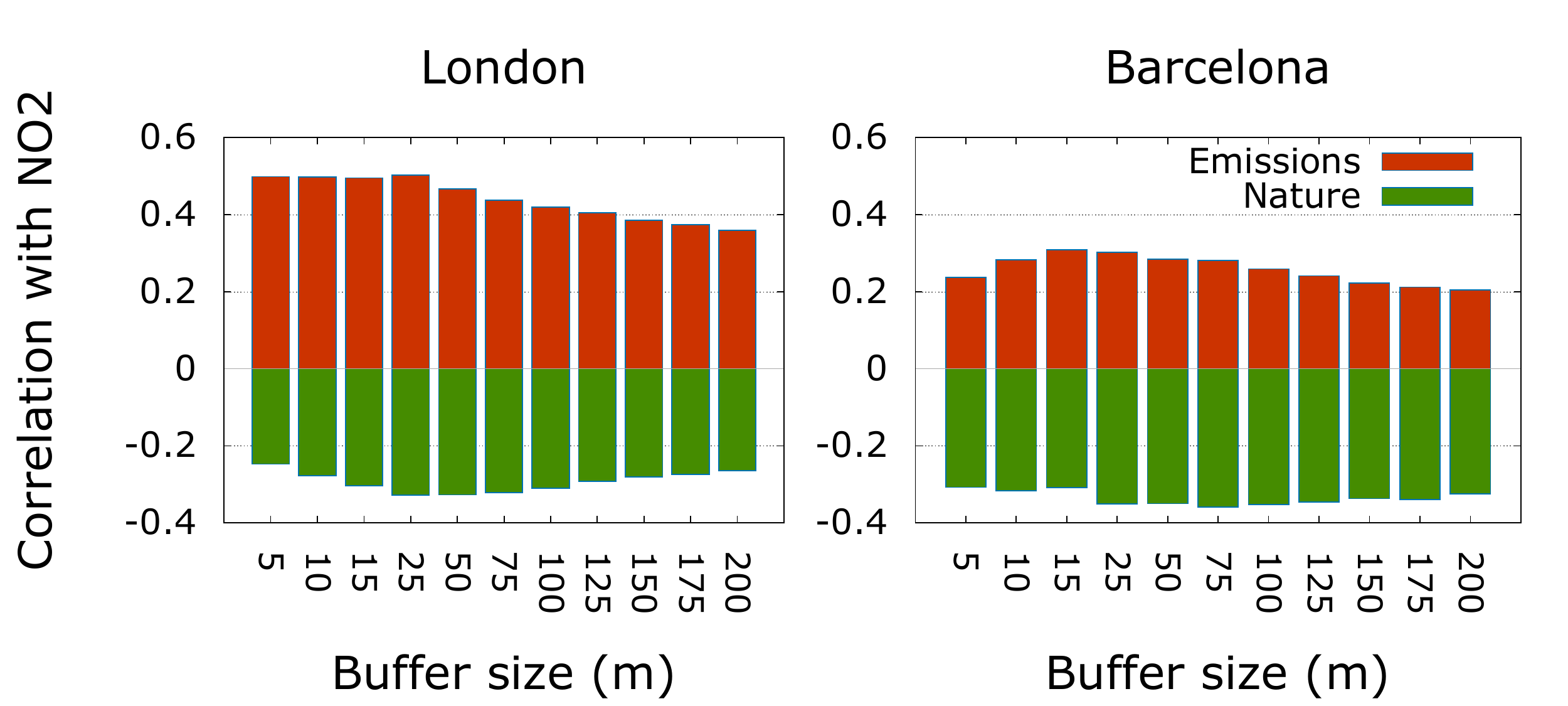}
\caption{Correlation with $NO_2$ vs. buffer size}
\label{fig:corr_vs_buffersize}
\vspace{-5mm}
\end{center}
\end{figure}

\begin{figure*}[t!]
    \subfloat[London, nature\label{fig:map_a}]{%
     \includegraphics[width = 0.32\textwidth]{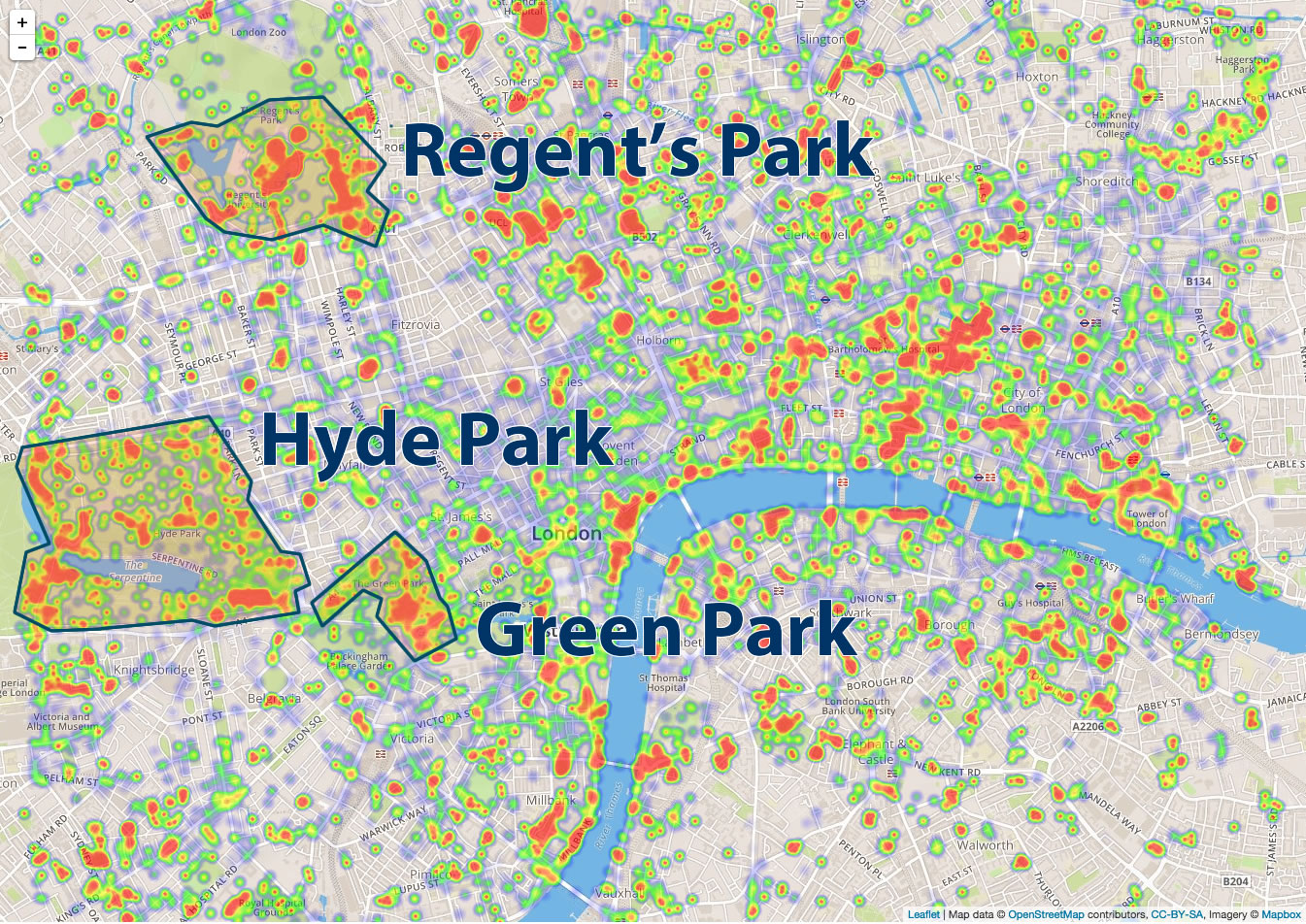}
    }
    \hfill
    \subfloat[London, emissions\label{fig:map_b}]{%
   \includegraphics[width = 0.32\textwidth]{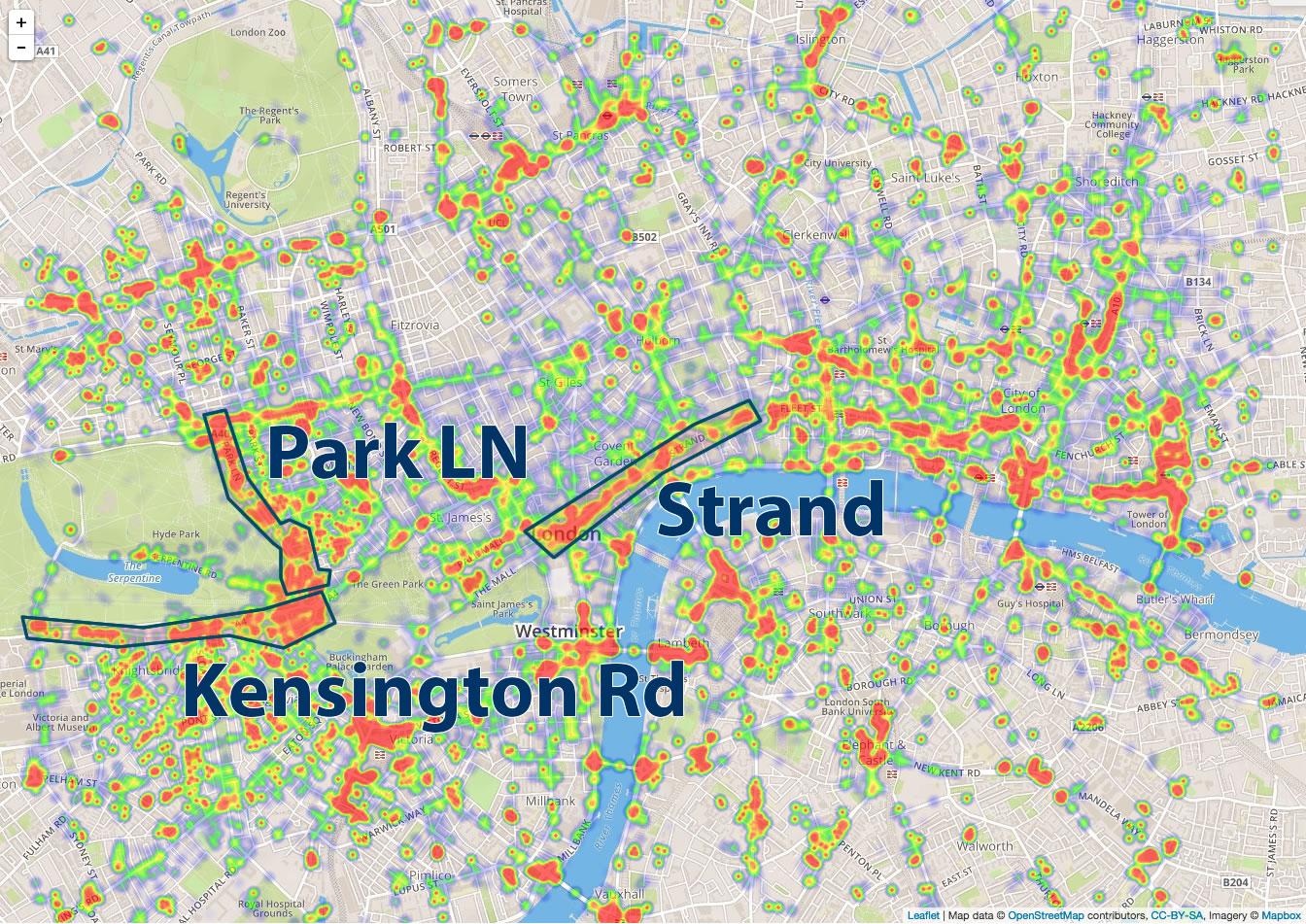}
    }
  \hfill
    \subfloat[London, animals\label{fig:map_c}]{%
       \includegraphics[width = 0.32\textwidth]{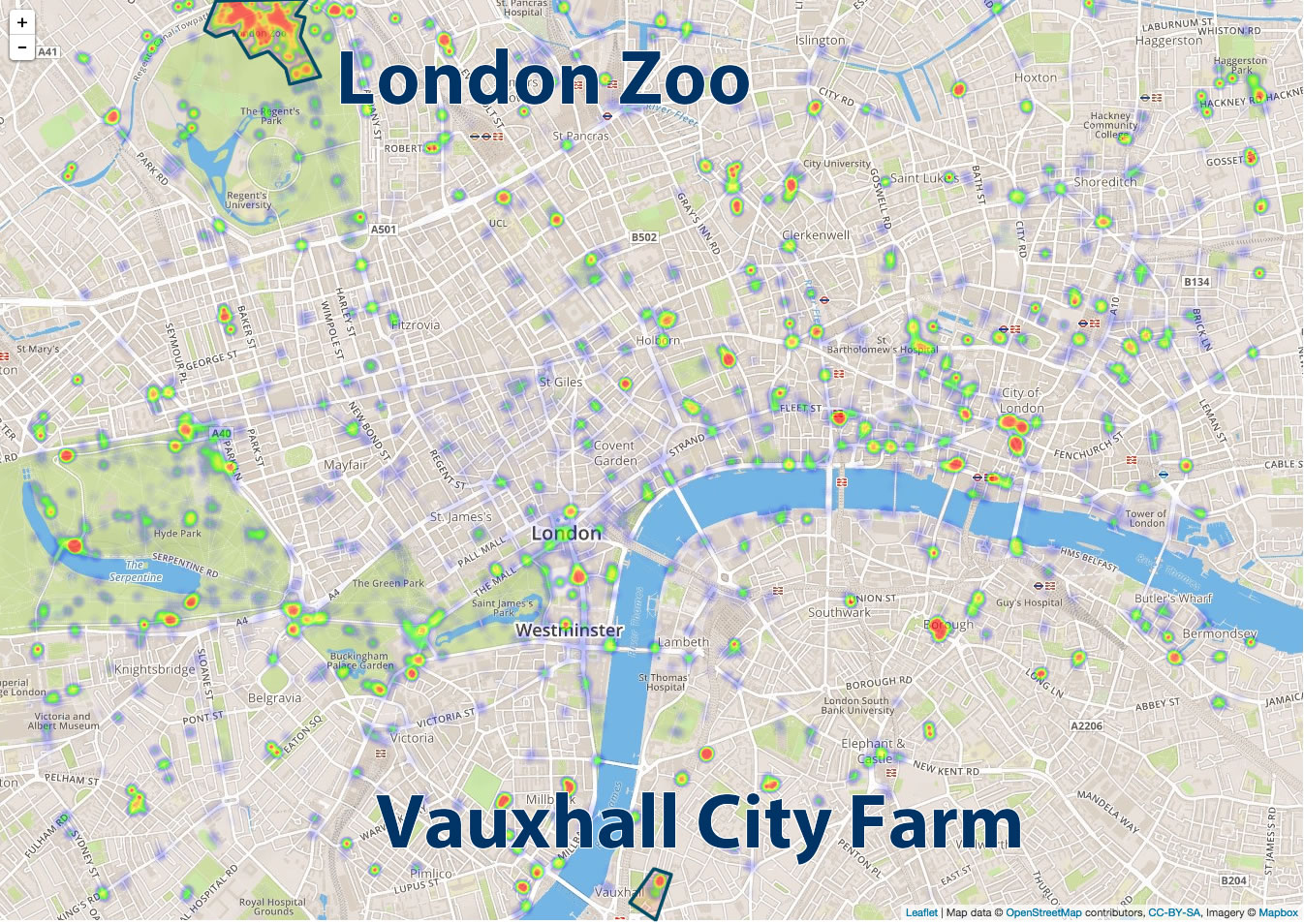}}
			\\
		\subfloat[Barcelona, nature\label{fig:map_d}]{%
     \includegraphics[width = 0.32\textwidth]{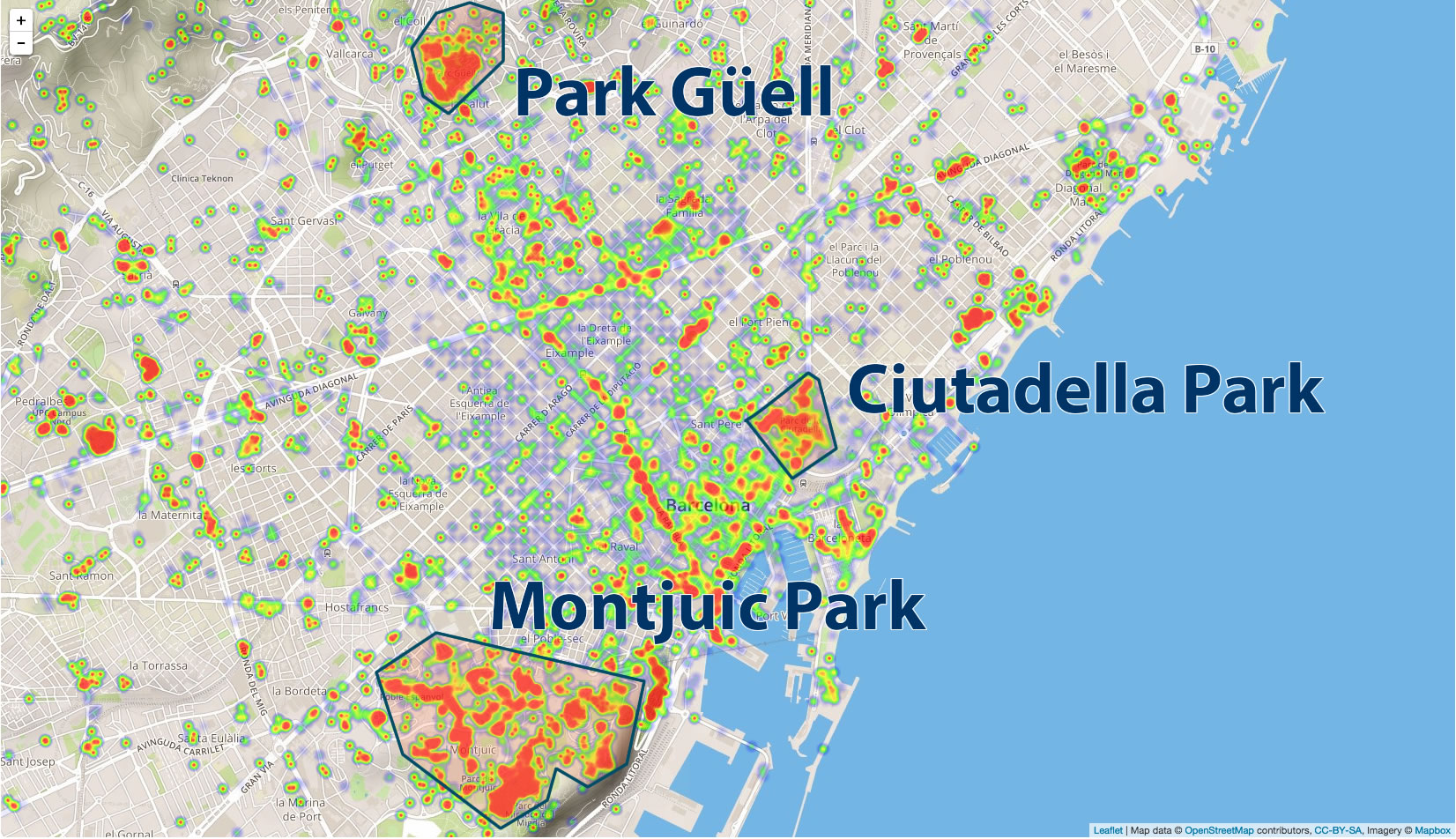}
    }
    \hfill
    \subfloat[Barcelona, emissions\label{fig:map_e}]{%
   \includegraphics[width = 0.32\textwidth]{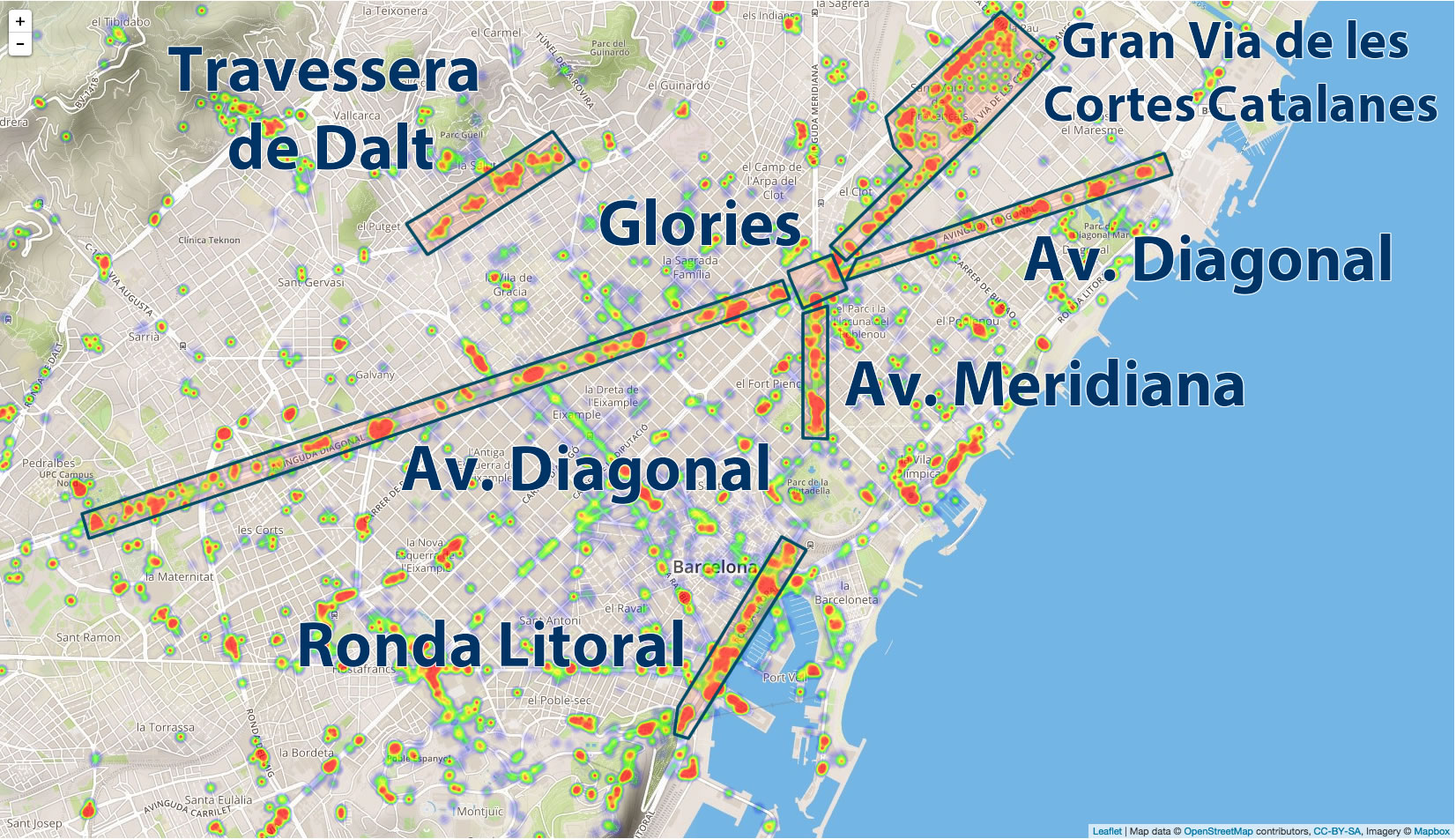}\\
    }
  \hfill
    \subfloat[Barcelona, animals\label{fig:map_f}]{%
       \includegraphics[width = 0.32\textwidth]{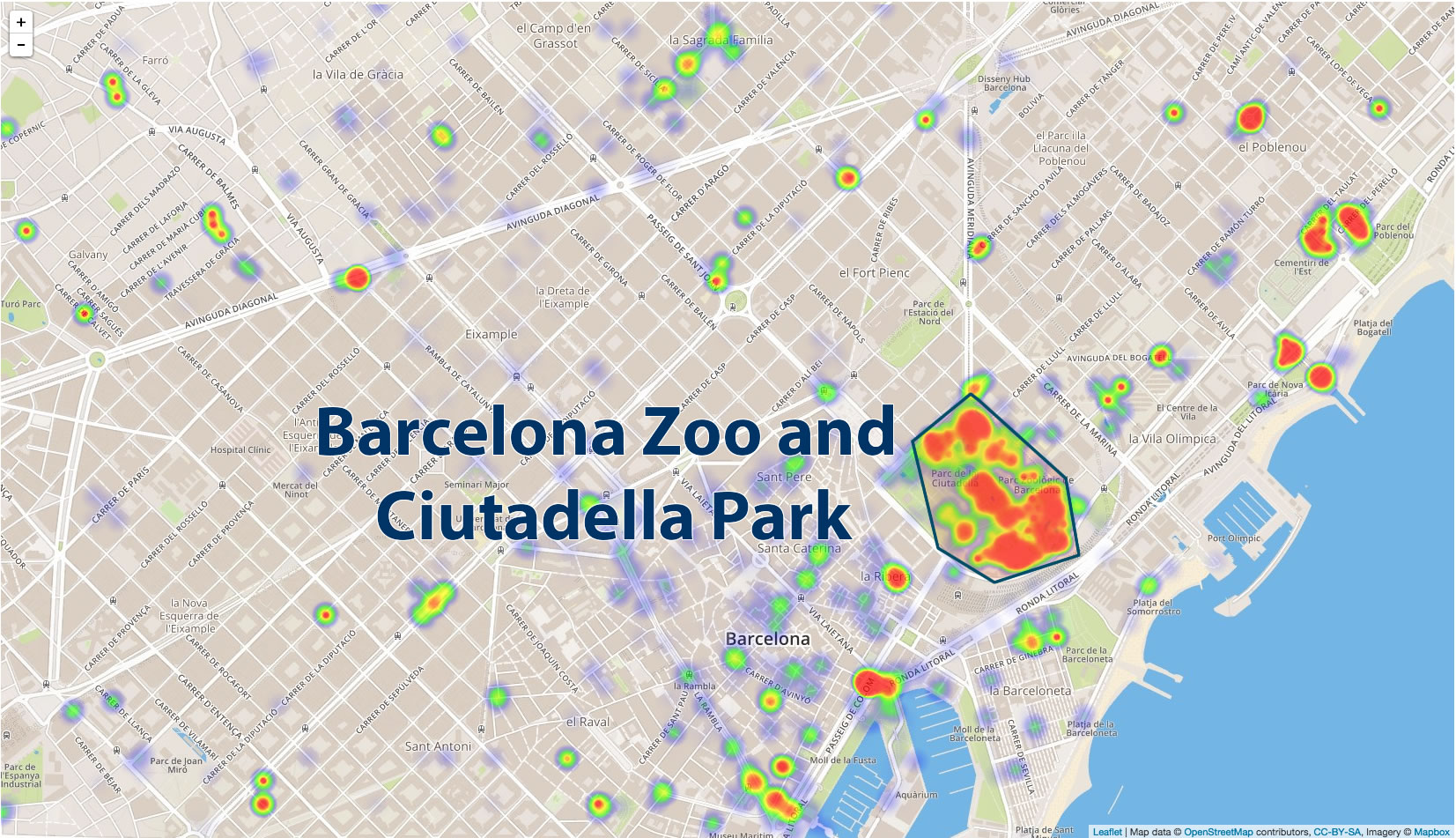}\\
    }
\caption{Heatmaps of smell-related tag intensity}
\vspace{-1em}
\label{fig:maps_basenotes}
\end{figure*}

\mbox{} \\
\textbf{Q2.} \textit{To which extent the smell categories of \textit{emissions} and \textit{nature} correlate with the air quality indicators, if the data comes from different social media sites?}  Across the three sites of Flickr, Instagram and Twitter (Figure~\ref{fig:correlations}), the correlations of pollutant concentrations are consistently positive with \emph{emissions} (red bars) and negative with \emph{nature} (green bars). The correlations are lower for Instagram and Twitter, slightly in London and moderately in Barcelona. That might be explained by three main reasons. First, the smaller the dataset, the lower the correlations (and Twitter is the smallest dataset among the three, as Table~\ref{tab:datasets} showed). Second, Twitter is less geographically salient than Flickr, as it has been previously shown~\cite{quercia13psycho}. Third, the location errors might differ across users of different services. As this might be yet another factor contributing to those differences in the correlation results, we test it next. 

\mbox{ } \\
\textbf{Q3.} \textit{Do our correlations depend on the size of the street buffer?} 
They do but not to a great extent. As the size increases, the correlations slightly degrade (Figure~\ref{fig:corr_vs_buffersize}). Therefore, if the buffer is too large, then matched tags are only loosely related to what is actually happening on the street. By contrast, if the buffer is too small (say below 20 meters), the spatial unit of study is excessively restricted, resulting into data sparsity problems. For both cities, a buffer size of 25 meters best strike a good balance between having relevant data and avoiding sparsity. 

\mbox{} \\
After having analyzed the relationship between pollutants and smell categories, we test whether, by mapping those categories, we are able to confirm what we have found so far. For an easier representation on heatmaps, we transform our smell and air quality vectors into the corresponding $z$-scores. Street segments with zero values are those experiencing the city's average presence of a smell category or of a pollutant. Those segments with values below (above) zero are  experiencing conditions below (above) the city's average. As one expects, the \textit{nature} category is present where the emissions category is absent, and vice-versa (Figure~\ref{fig:maps_basenotes}). In London, Hyde Park experiences high levels in the \emph{nature} category, and, conversely, the trafficked streets at its boundaries experience high levels in the \emph{emissions} category. In Barcelona, the same goes for Montjuic Park and Park Guell, and for the nearby streets of Ronda Litoral and the Travessera de Dalt. Finally, the last column of Figure~\ref{fig:maps_basenotes} shows the heatmaps for \emph{animal} smells, that is registered around both Barcelona's and London's zoos. In London, other smaller animal small holdings (e.g., Vauxhall City Farm) also emerge.

\subsection{Mid-level Notes of Urban Smell}

\begin{figure}[t!]
\begin{center}
\includegraphics[width=0.9\columnwidth]{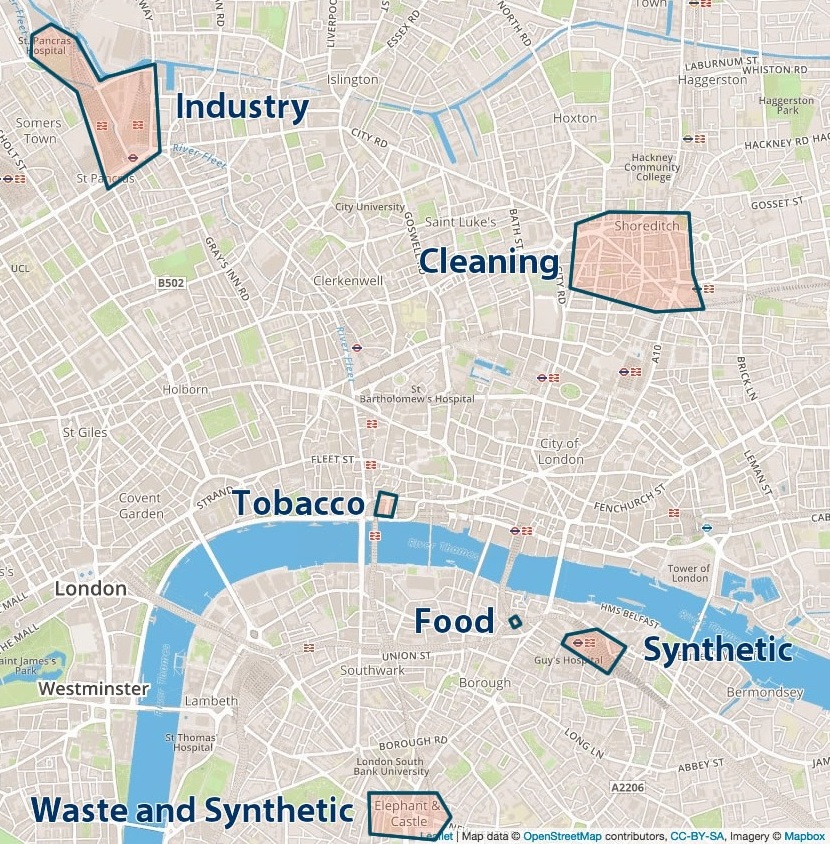}
\caption{Mid-level notes of urban smell for London}
\label{fig:map_midlevel_ldn}
\vspace{-5mm}
\end{center}
\end{figure}

\begin{figure}[t!]
\begin{center}
\includegraphics[width=0.9\columnwidth]{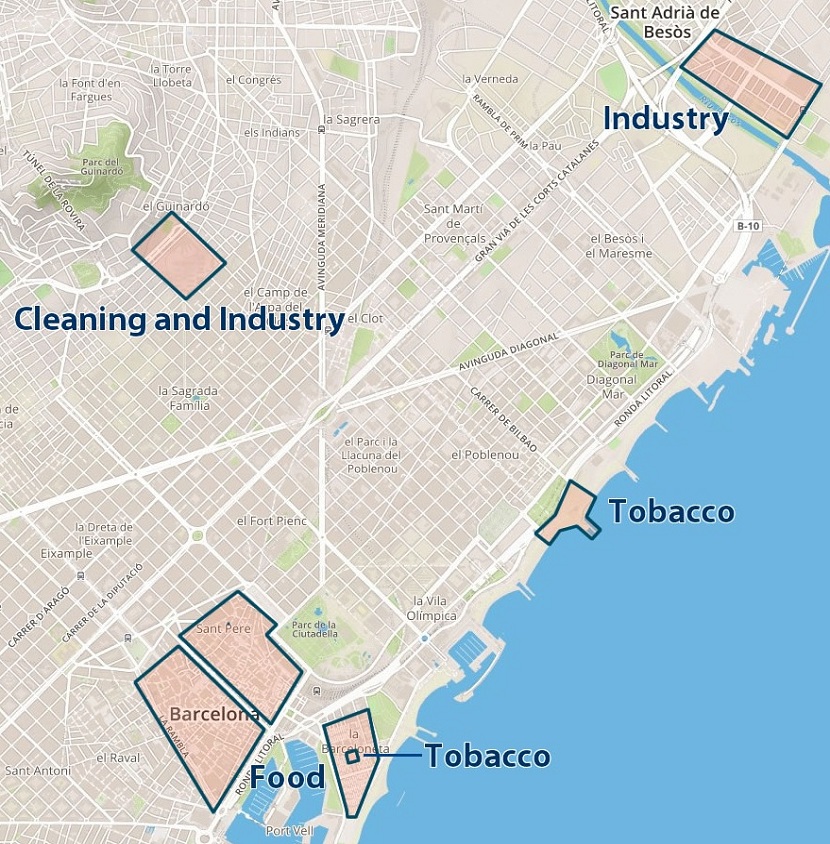}
\caption{Mid-level notes of urban smell for Barcelona}
\label{fig:map_midlevel_bcn}
\vspace{-5mm}
\end{center}
\end{figure}

With social media, we have seen that we are able to capture background smells (base-level smell notes). Now we will show that we are also able to capture smells localized in specific areas (mid-level smell notes). Consider the maps showing the presence of some of the remaining smell categories (other than the three we have discussed) in London (Figure~\ref{fig:map_midlevel_ldn}) and Barcelona (Figure~\ref{fig:map_midlevel_bcn}). Those maps suggest that the remaining smell categories are not dominant but are localized in specific areas:
\begin{description}
\item Smells of \emph{food} are localized around food markets (Boqueria market, Barcelona; Borough Market, London) and in areas where restaurants tend to cluster (Born and Barceloneta, Barcelona).

\item Smells of \emph{waste} and \emph{smoking} are found in areas enjoying the evening economy (Barceloneta, and Bogatell beach, Barcelona; Blackfriars, and Elephant and Castle, London). That is partly because, the encouragement of the evening economy has increased the levels of waste in city streets (e.g., odors of urine, and cigarette)~\cite{victoria2013}, in particular following the criminalization of smoking in enclosed public places.

\item Strong presence of  \emph{cleaning} (product) smells are detected in Shoreditch, London. More expectedly, those smells plus \emph{chemical} ones are detected around big industrial facilities (Sant Adria, Barcelona), hospitals (Hospital San Pau, Barcelona), and big railways stations (King's Cross, London). 
\end{description}

\section{Discussion}
\label{sec:discussion}

This work aims to engage with academics and built environment professionals who are passionate about the multisensory experience of cities. It highlights the positive role that `smell' as opposed to `air pollution' can play in the environmental experience. Next, we discuss some of the limitations of our work and some of the opportunities it opens up.

\subsection{Limitations}

The way people perceive odors is individually, socially, and contextually situated creating a nuanced dataset.

\mbox{ } \\
\textbf{Individual factors.} Individual factors. Personal characteristics affect smell perceptions. Females have higher olfactory performance than males. Age has a limiting influence on smell performance, with 50\% of people experiencing a major loss in olfactory function over the age of 65. Finally, smoking habits reduce smell performance~\cite{vennemann08}.

\mbox{ } \\
\textbf{Socio-cultural factors.} Urban odor classification is an attempt to model the range of background and episodic odors detected and reported. However, it is not an exhaustive listing: cities in parts of the world with extreme climates, such as high humidity or sub-zero temperatures. are likely to be characterized by odors not identified in our northern European-based classification. Having said that, we should stress that most of the smell groups in the classification would be present in the vast majority of cities. Afterall, where there are densely populated areas, there will always be food, waste and materials.

\mbox{ } \\
\textbf{Contextual factors.} Urban planning and the resulting city layout have significant impact on odor detection in the city. The grid layout of New York City, for example, encourages large-scale collective odor experiences as it was designed in a way to facilitate airflow using prevailing westerly winds to dissipate the disease-carrying miasmas of the late 18th century. In October 2005, a sweet, sirupy odor was detected across the city. The smell was pleasant (i.e., a combination of maple syrup and caramel), yet it resulted into hundreds of calls to the city's emergency services. `The aroma not only revived memories of childhood, but in a city scared by terrorism, it raised vague worries about an attack deviously cloaked in the smell of grandma's kitchen'~\cite{depalma05}. Also, long-distance smell detection is highly temporal, dependent on weather conditions, wind patterns, and seasonal waves of activity, with air temperature directly influencing odor strength and volatility. Finally, in a twenty-four-hour city, the same street will host different activities at different times: activities associated with, for example, the cafe and retail culture during the daytime, and drinking culture in the evening.

\subsection{Opportunities}

Despite these limitations, our work offers new ways of facilitating olfactory interpretations of places for a variety of disciplines.

\mbox{ } \\
\textbf{Urban Planning.} One hundred sites In Japan have been declared as protected because of their `good fragrance'. However, the general situation in the rest of the world greatly differs: urban planners to date have tended to think about smells in terms of management of bad odors rarely considering preserving and celebrating the smells that people like. There are a number of ways that the urban smellscape can be altered; manipulating the air flow by changing the street layout, pedestrianization to alter traffic emissions (categories from Figure~\ref{fig:taxonomy} can be mapped onto cities to add weight to arguments to reduce emissions), the creation of restorative environments through the planting of trees, greenspaces and waterways (categories from Figure~\ref{fig:taxonomy} can be mapped over a variety of cities to depict olfactory perception of green spaces), and the strategic placement of car stopping points are just a few examples. City officials do not fully consider the opportunities presented by the sense of smell simply because they have been the victims of a discipline's negative perspective. We hope that our work might help them re-think their approaches and use olfactory opportunities to create stimulating multi-sensory places.

\mbox{ } \\
\textbf{Computer Science.} In the near future, new way-finding tools might well suggest not only shortest routes between points but also short ones that are olfactorily pleasant (e.g. runners might wish to avoid emission-infused streets). Our methodology allows for the development of new tools to map urban smellscapes.

\mbox{ } \\
\textbf{Arts \& Humanities.} Contemporary art, design and philosophy tend towards a phenomenological understanding, using our senses to constantly rediscover the world we live in. In this vein, one of the co-authors collects and analyses olfactory data derived from smellwalks, visualizing the scents and their possible locations in the city using a variety of creative mapping practices. Her work exhibits internationally incorporating data visualisation of the smellscape and a variety of synthetically and naturally made scents. Olfactory artists are likely to profit from our methodology, incorporating a wider range of digital traces in their production of artwork.

\mbox{ } \\
\textbf{Public Engagement.} In addition to academic research, the general public might also benefit by contributing to the development of a critical voice for the positive and negative role that smell has to play in the city. 

\mbox{ }
The urban smellscape is a complex set of sensorial fragments, and it is debated as to whether a smellscape can ever be fully known. For example, when we consider a landscape it forms a continuous, integrated and defined space whereas the smellscape is a dynamic and fluid entity. It is impossible for a number of humans to detect the entire smellscape of an area as a whole at any one point in time. Smellwalks only partly solve those problems since they suffer from two main biases: \textit{a)} sample bias (participants are not representative of the general population); and \textit{b)} response bias (people might perform tasks in the walk differently than how they would ‘in the wild’ because of the Hawthorne effect~\cite{haweff}). Social media partly reduces both biases, in that, representative user samples can be extracted (reduced sample bias), and data can be captured unobtrusively (with lack of experimental demands, the response bias will be limited). In addition to being unobtrusive, social media appears insightful for capturing elements of urban smellscapes: our results suggest that it is possible to effectively track urban odors from digital traces when combined with smell-related words learned from smellwalks. This result should come as no surprise to practitioners of mixed methods research.

\section{Conclusion}

We have contributed to the growing body of literature on how people sensually experience the city. 
There has been research on how we see the city and on how we hear the city, but not much on how we smell the city. This work is the first in examining the role of social media in mapping urban smell environments in an unobtrusive way. We hope to empower designers, researchers, city managers by offering them a number of methodological tools and practical insights to re-think the role of smell in their work.

In the future, we would like to conduct a more comprehensive multi-sensory evaluation by  exploring how  the sound, visual, and olfactory aesthetics compare in the same city. We are also interested in capturing fleeting odors. As these are localized in space and time (the high notes of the urban smellscape), they  cannot be easily identified on social media and might require the design of new mobile apps to facilitate crowdsourcing their collection.

\small
\balance
\bibliographystyle{aaai}
\bibliography{bibliography}

\end{document}